\documentclass[12pt]{article}

\begin{document}
\newenvironment{Eqnarray}%
         {\arraycolsep 0.14em\begin{eqnarray}}{\end{eqnarray}}

\def\slash#1{{\rlap{$#1$} \thinspace/}}

\hspace{1cm}

\begin{flushright} 
November, 2000 \\
KEK-TH-723 \\
TIT/HEP-458 \\
\end{flushright} 

\vspace{0.1cm}

\begin{Large}
       \vspace{1cm}
  \begin{center}
   {Supercurrent Interactions in \\
 Noncommutative Yang-Mills and IIB Matrix Model}      \\
  \end{center}
 \end{Large}

  \vspace{1cm}

\begin{center}
           YUSUKE KIMURA $^{a)}$$^{b)}$\footnote{
e-mail address : kimuray@post.kek.jp }
and YOSHIHISA KITAZAWA $^{a)}$\footnote{
e-mail address : kitazawa@post.kek.jp} \\
        
       \vspace{0.7cm} 
        $a)${\it  High Energy Accelerator Research Organization (KEK),}\\
               {\it Tsukuba, Ibaraki 305-0801, Japan} \\
       \vspace{0.3cm}   
        $b)$ {\it  Department of Physics, Tokyo Institute of Technology, \\
 Oh-okayama, Meguro-ku, Tokyo 152, Japan}\\
\end{center}
        \vspace{0.8cm}

\begin{abstract}
\noindent
\end{abstract}

It is known that noncommutative Yang-Mills is equivalent to 
IIB matrix model with a noncommutative background, 
which is interpreted as a twisted reduced model. 
In noncommutative Yang-Mills, long range interactions 
can be seen in nonplanar diagrams after integrating high momentum modes. 
These interactions can be understood as block-block interactions 
in the matrix model. 
Using this relation, we consider long range interactions in noncommutative 
Yang-Mills associated with fermionic backgrounds. 
Exchanges of gravitinos, 
which couple to a supersymmetry current, 
are examined.

\newpage

\section{Introduction}

\hspace{0.4cm}
Several kinds of Matrix Model 
have been proposed\cite{BFSS,IKKT} 
to study the nonperturbative aspects of string theory or M theory.
These proposals are based on the 
developments of D-brane physics. 
D-branes have been shown to play a fundamental role 
in nonperturbative string theory\cite{pol,Taylor}. 
A notable point is that 
supersymmetric gauge theory can be obtained 
on their world-volume 
as their low energy effective theory. 
The idea of matrix models is that 
supersymmetric gauge theory can  
describe string or M theory. 

IIB Matrix Model is one of these proposals\cite{IKKT}.
It is  a large $N$ reduced model\cite{reduced} 
of ten-dimensional supersymmetric 
Yang-Mills theory and 
the action has a matrix regularized form of 
the Green-Schwarz action of IIB superstring. 
It is postulated that it gives the constructive definition of 
type IIB superstring theory. 
This model 
has ten dimensional ${\cal N}=2$ supersymmetry, 
which implies the existence of gravitons.
In the matrix model, gravitational interactions arise as 
quantum effects. 
In fact, the leading long range interaction 
in the matrix model 
is identified with the supergraviry results. 
Gravitons couple to energy-momentum tensor 
and the interaction 
between the separate objects  
exhibits the graviton exchange process\cite{IKKT}. 
We expect that IIB matrix model 
should reproduce the interactions 
which are mediated by the whole 
multiplet in IIB supergravity.
In \cite{IKKT}, one-loop effective action is calculated only 
for bosonic backgrounds. 
By considering fermionic backgrounds\cite{SuyaTsu,TaylorRaam}, 
fermionic particles such as gravitinos or dilatinos 
are expected to be seen in the matrix model calculation. 
Hence it is important to consider fermionic backgrounds 
to check that the IIB matrix model can reproduce 
the interactions expected in IIB supergravity.

Recently, noncommutative Yang-Mills theories have been 
studied in many situations. 
It first appeared within the framework 
of toroidal compactification of Matrix theory\cite{CDS}. 
It is discussed in \cite{SW} that 
the world volume theory on D-branes 
with NS-NS two-form background 
is described by noncommutative Yang-Mills theory. 
In matrix models, space-time coordinates are represented  
by matrices. Therefore the noncommutativity appears naturally 
and matrix models are considered to be closely 
related to noncommutative geometry. 
It was shown\cite{AIIKKT,BM,AMNS} that in the matrix model picture 
noncommutative Yang-Mills theory 
is equivalent to twisted reduced models\cite{GAO}. 
Twisted reduced models are obtained by 
expanding the model around noncommutative backgrounds. 
A noncommutative background is a D-brane-like background 
which is a solution of equation of motion and 
preserves a part of supersymmetry. 
It is well known that gauge theory is realized 
in the world-volume of D-branes as their low energy effective theory. 
In IIB matrix model, 
gauge theory is realized as twisted reduced models. 
Noncommutative Yang-Mills as a twisted reduced model 
has been studied in \cite{IIKK,IKK,IIKK4038,12258,AMNS2}.

Noncommutative field theory has a lot of interesting 
properties which are absent in ordinary field theory. 
While amplitudes for planar diagrams in the noncommutative theory are 
the same as those in the commutative theory 
up to a phase factor associated with external lines, 
amplitudes for nonplanar diagrams in the noncommutative theory 
are ultraviolet finite due to the oscillation 
of the phase factor\cite{GAO,filk,BiSu,IIKK}. 
Perturbative dynamics of noncommutative field theory 
has been further studied in \cite{MRS,Haya} 
and it is pointed out that the effective action 
has infrared singular behavior in nonplanar diagrams. 
After integrating high momentum modes, 
long range interactions, which are absent in ordinary field theories, 
can be obtained. 
This infrared singular behavior 
may be related with the 
propagation of massless particles in the bulk. 
This behavior reminds us of the channel duality in string theory. 
High momentum modes at open string one loop level on the brane
corresponds to 
the exchange of low momentum modes in closed string, 
which propagates in the bulk. 
A nonplanar one-loop diagram is topologically equivalent 
to a tree level diagram in closed string theory. 
This interaction can be understood as block-block 
interactions in the matrix model picture\cite{IKK}. 
These long range interactions, or the propagation 
of massless particles, are universal property 
of noncommutative field theories and the matrix model. 

In this paper, we consider supercurrent interactions 
in noncommutative Yang-Mills and IIB matrix model 
using the formulation of noncommutative Yang-Mills 
as twisted reduced models. 
We find that 
block-block interactions at order $1/r^{8}$ 
in the matrix model with fermionic background 
give gravitino exchange processes at order $1/r^{9}$ 
in noncommutative Yang-Mills. 
The interaction which decays as $1/r^{9}$ is interpreted 
as the propagation of a massless fermion in ten dimensions 
and does not depend on the information of the extension 
of the matrix eigenvalues. 
Then it is presented that 
one of the gravitinos in IIB supergravity 
couples to a supersymmetry current
which is a Noether current associated 
with supersymmetry in IIB matrix model. 
The organization of this paper is as follows. 
In section 2, we review IIB matrix model and 
its relation to noncommutative Yang-Mills theory. 
In the matrix model picture, noncommutative Yang-Mills 
is equivalent to twisted reduced model. 
In section 3, we consider the long range interactions 
with fermionic backgrounds in IIB matrix model and 
noncommutative Yang-Mills. 
These long range interactions arise from nonplanar diagrams in 
noncommutative Yang-Mills. 
Long range interactions in noncommutative Yang-Mills 
which decay as $1/r^{9}$ are obtained. 
It is shown that this interactions are due to the gravitino 
exchange and this gravitino couples to a supersymmetry current 
in section 4. 
The interactions between supersymmetry currents 
via a gravitino exchange in ten-dimensional supergravity 
are computed. 
Then we compare the matrix model calculation 
with a supergravity calculation. 
Section 5 is devoted to conclusions and discussions.


\section{IIB matrix model and noncommutative Yang-Mills}

\hspace{0.4cm}
In this section, we review 
IIB matrix model\cite{IKKT,AIKKTT} 
and its relation to noncommutative Yang-Mills\cite{AIIKKT,IIKK,IKK}.

We begin with the action 
which is defined by the following form: 

\begin{equation}
 S= -\frac{1}{g^{2}} Tr \left( \frac{1}{4} \left[ A_{\mu} ,A_{\nu}\right] 
         \left[A^{\mu} ,A^{\nu} \right]  
  +\frac{1}{2}\bar{\psi } \Gamma^{\mu} \left[ A_{\mu},\psi \right] \right) . 
 \label{action}
\end{equation}

\noindent 
Here $\psi$ is a ten dimensional Majorana-Weyl spinor field, 
and $A_{\mu}$ and $\psi$ are $N \times N$ hermitian matrices. 
This model is the large $N$ reduced model of ten-dimensional 
$\cal{N}$=1 $U(N)$ supersymmetric Yang-Mills theory. 
This is based on the observation that 
in the large $N$ t'Hooft limit $U(N)$ gauge theory is equivalent 
to its reduced model which is obtained by reducing the 
space-time volume to a single point\cite{reduced}.

This model has the manifest ten dimensional Lorentz symmetry and 
following symmetries:

\begin{Eqnarray}
\delta^{(1)} \psi &=& \frac{i}{2} 
  \left[ A_{\mu} ,A_{\nu}\right] \Gamma^{\mu\nu} \epsilon,  \cr
\delta^{(1)} A_{\mu} &=& i\bar{\epsilon}
     \Gamma^{\mu}\psi,  \label{susy1}
\end{Eqnarray}
and
\begin{Eqnarray}
\delta^{(2)} \psi &=& \xi,  \cr
\delta^{(2)} A_{\mu} &=& 0, \label{susy2}
\end{Eqnarray}

\noindent and translation symmetry:

\begin{equation}
\delta A_{\mu} = c_{\mu} {\bf 1}.
\end{equation}

\noindent If we interpret the eigenvalues of $A_{\mu}$ 
as the space-time coordinates, 
we can regard the above symmetry as ${\cal N}=2$ supersymmetry\cite{IKKT}. 
We take a linear combination of $\delta^{(1)}$ and $\delta^{(2)}$ as 

\begin{Eqnarray}
\tilde{\delta}^{(1)} &=& \delta^{(1)} + \delta^{(2)},   \cr
\tilde{\delta}^{(2)} &=& 
   i\left(\delta^{(1)} - \delta^{(2)} \right) .
\end{Eqnarray}

\noindent 
We can obtain ${\cal N}=2$ supersymmetry algebra:

\begin{Eqnarray}
\left(
\tilde{\delta}_{\epsilon}^{(i)}\tilde{\delta}_{\xi}^{(j)}
 -\tilde{\delta}_{\xi}^{(j)}\tilde{\delta}_{\epsilon}^{(i)}
\right) \psi &=& 0, \cr
\left(
\tilde{\delta}_{\epsilon}^{(i)}\tilde{\delta}_{\xi}^{(j)}
 -\tilde{\delta}_{\xi}^{(j)}\tilde{\delta}_{\epsilon}^{(i)}
\right) A_{\mu} &=& 2i \bar{\epsilon} \Gamma^{\mu} \xi \delta_{ij}.
\end{Eqnarray}

\noindent 
The classical equations of motion of (\ref{action}) are 

\begin{equation}
\Gamma_{\mu}[A_{\mu},\psi]=0,  \label{eom1}
\end{equation}

\begin{equation}
[A_{\mu},[A_{\mu},A_{\nu}]]=-\bar{\psi}\Gamma_{\nu}\psi . 
\label{eom2}
\end{equation}

\vspace{0.6cm}
In the latter part of this section, we briefly review 
the formulation\cite{AIIKKT} of 
noncommutative gauge theory as twisted reduced models.
We expand the theory around the following classical solution, 

\begin{equation}
\left[ \hat{p}^{\mu} , \hat{p}^{\nu} \right] = i B^{\mu \nu}, 
\label{classicalsol} 
\end{equation} 

\noindent where $B_{\mu \nu}$ is  anti-symmetric tensor 
and proportional to a unit matrix. 
This is a solution of (\ref{eom2}) with $\psi=0$ and 
corresponds to a BPS background 
($\xi =\pm 1/2B^{\mu \nu}\epsilon$)
\cite{IKKT}.  
We assume the rank of $B_{\mu \nu}$ to be $d$ 
and define its inverse $C^{\mu \nu}$ 
in $d$ dimensional subspace. 
$\hat{p}^{\mu}$ satisfy the canonical commutation relations 
and span the $d$ dimensional phase space. 
The volume of the phase space is 
$V_{p} = n(2\pi)^{d/2} \sqrt{\det B}$. 
Then we expand $A_{\mu}=\hat{p}_{\mu}+\hat{a}_{\mu}$ and 
Fourier-decompose $\hat{a}_{\mu}$ and $\hat{\psi}$ as 

\begin{equation}
\hat{a}_{\mu}=\sum _{k} \tilde{a}_{\mu}(k) 
\exp(i C^{\mu \nu} k_{\mu} \hat{p}^{\nu}),
\end{equation}

\begin{equation}
\hat{\psi}=\sum _{k} \tilde{\psi}(k) 
\exp(i C^{\mu \nu} k_{\mu} \hat{p}^{\nu}).
\end{equation}
$\exp(i C^{\mu \nu} k_{\mu} \hat{p}^{\nu})$ is the eigenstate 
of $P_{\mu}=[\hat{p}^{\mu}, \cdot  ]$  with eigenvalue $k_{\mu}$.
The Hermiticity requires that 
$\tilde{a}_{\mu}^{\ast}(k) = \tilde{a}_{\mu}(-k)$
and $\tilde{\psi}_{\mu}^{\ast}(k) =\tilde{\psi}_{\mu}(-k)$. 
Let $\Lambda$ be the extension of each $\hat{p}_{\mu}$.
The volume of one quantum in this phase space is 
$\Lambda^{d} /N=\lambda^{d}$
where $\lambda$ is the spacing of the quanta, 
say, noncommutative scale. 
$B$ , which is the component of $B_{\mu\nu}$, 
is related to $\lambda$ as $B=\lambda^{2}/2\pi$.
$k^{\mu}$ is quantized 
in the unit of $k_{\mu}^{min}=\Lambda/N^{2/d}=\lambda/N^{1/d}$.
The range of $k_{\mu}$ is restricted as 
$-N^{1/d} \lambda/2 \le k_{\mu} 
\le N^{1/d} \lambda/2.$

Consider the map from a matrix to a function as 

\begin{equation}
\hat{a}_{\mu} \rightarrow 
a_{\mu}(x)=\sum _{k} \tilde{a}_{\mu}(k) 
\exp(i k_{\mu} x^{\mu}).  
\end{equation}
\begin{equation}
\hat{\psi} \rightarrow 
\theta(x)=\sum _{k} \tilde{\theta}(k) 
\exp(i k_{\mu} x^{\mu}).  \label{map}
\end{equation}

\noindent We consider this field as the gauge field in 
noncommutative gauge theory.
Under this map, we obtain the following map, 

\begin{equation}
\hat{a}\hat{b} \rightarrow 
a(x) \star b(x), 
\end{equation}

\noindent 
where $\star$ is the star product defined as follows, 

\begin{equation} 
a(x) \star b(x) \equiv 
\exp \left( \frac{i C^{\mu\nu}}{2} 
\frac{\partial^2}{\partial \xi^{\mu} \partial \eta^{\nu}} \right)
a(x+\xi)b(x+\eta) \mid _{\xi=\eta=0}.    \label{star}
\end{equation}

\noindent 
$Tr$ over matrices can be mapped 
on the integration over functions as 

\begin{equation}
Tr[\hat{a}] =\sqrt{\det B} 
\left(\frac{1}{2\pi} \right)^{\frac{d}{2}}
\int d^{d}x a(x) .
\end{equation}

\noindent 
Using these rules, 
the adjoint operator of $\hat{p}_{\mu}+\hat{a}_{\mu}$ is mapped 
to the covariant derivative:

\begin{equation}
\left[\hat{p}_{\mu} + \hat{a}_{\mu},\hat{o} \right] 
\rightarrow \frac{1}{i} \partial_{\mu} o(x) 
+a_{\mu}(x) \star o(x) - o(x) \star a_{\mu}(x)
\equiv \frac{1}{i}\left[D_{\mu},o(x) \right]_{\star},   \label{covder}
\end{equation}
\noindent and 
\begin{equation}
f_{\mu\nu}=i[A_{\mu},A_{\nu}] \rightarrow
-B_{\mu\nu}
+\partial_{\mu}a_{\nu}-\partial_{\nu}a_{\mu}+i[a_{\mu},a_{\nu}]_{\star}. 
\label{fieldst}
\end{equation}

\noindent 
The equations of motion (\ref{eom1}) and (\ref{eom2}) 
of the matrix model are mapped to 

\begin{equation}
\Gamma_{\mu}[D_{\mu},\theta]_{\star}=0, 
\end{equation}
\begin{equation}
[D_{\mu},f_{\mu \nu}]_{\star}=(\Gamma_{\nu})_{\alpha\beta}
\bar{\theta}_{\alpha}\star\theta_{\beta}.  \label{eomncym}
\end{equation}

\noindent 
By applying these rules to the action (\ref{action}), 
$U(1)$ noncommutative Yang-Mills theory has been obtained: 

\begin{Eqnarray}
&&-\frac{1}{4g^{2}} Tr \left[ A_{\mu} ,A_{\nu}\right] 
         \left[A^{\mu} ,A^{\nu} \right]  \cr
&\rightarrow& 
\frac{d N B^2}{4g^2}
-\sqrt{\det B}(\frac{1}{2\pi})^{d\over 2}
\int d^{d}x
\frac{1}{g^2} (\frac{1}{4}
[D_{\alpha},D_{\beta}][D_{\alpha},D_{\beta}]  \cr
&&-\frac{1}{2}[D_{\alpha},\phi_{a}][D_{\alpha},\phi_{a}]
+\frac{1}{4}[\phi_{a},\phi_{b}]
[\phi_{a},\phi_{b}])_{\star}  , 
\end{Eqnarray}
\noindent and 
\begin{Eqnarray}
&&-\frac{1}{2g^{2}}\bar{\psi } 
\Gamma^{\mu} \left[ A_{\mu},\psi \right] \cr
&\rightarrow&
-\sqrt{\det B}(\frac{1}{2\pi})^{d \over 2}
\int d^{d}x \frac{1}{2g^2 i} 
(\bar{\theta}\Gamma_{\alpha}[D_{\alpha},\theta] 
+\bar{\theta}\Gamma_{a}[D_{a},\theta] )_{\star} .
\end{Eqnarray}

\noindent 
where the indices $\alpha$ and $\beta$ run 
over the directions parallel to the brane and 
the indices $a$ and $b$ over the directions 
transverse to the brane. In the transverse direction, 
$a_{a}$ has been replaced by a scalar field $\phi_{a}$.
Although we have discussed the momentum space, 
the coordinate space is also embedded in the matrices 
of twisted reduced model through the relation 
$\hat{x}^{\mu}=C^{\mu\nu}\hat{p}_{\nu}$.
This relation says that 
the coordinate space is related to the momentum space.
This relation is relevant to T-duality\cite{IIKK4038}.


\section{Long range interaction with fermionic backgrounds} 
\hspace{0.4cm} 
In this section, we consider quantum corrections 
of the matrix model. 
Computing the one-loop effective action between diagonal blocks, 
the gravitational interactions can be observed 
and IIB supergravity is expected to be reproduced. 
Graviton and dilaton exchange are examined in \cite{IKKT,IIKK4038}. 
In \cite{IKKT}, one-loop effective action is calculated 
without fermionic backgrounds. 
With fermionic backgrounds, gravitino and dilatino 
exchange processes are 
expected to be seen. 
One-loop effective action including fermionic backgrounds is 
examined in \cite{SuyaTsu} and 
in \cite{TaylorRaam} in BFSS matrix model. 
In the bosonic background, 
leading $1/r^{8}$ terms in IKKT model\cite{IKKT} and 
leading $1/r^{7}$ terms in BFSS model\cite{BFSS} 
are related by T-duality.

We now derive the one-loop effective action in fermionic 
backgrounds based on \cite{IKKT,SuyaTsu}. 
The matrices $A_{\mu}$ and $\psi$ 
are divided into  
the backgrounds and fluctuations:

\begin{equation}
A_{\mu}=p_{\mu}+a_{\mu} , 
\end{equation}
\begin{equation}
\psi = \theta +\varphi .
\end{equation}

\noindent 
The backgrounds have block-diagonal form:

\begin{equation}
A_{\mu}^{back}  
 =  \left( \begin{array}{c c c}
  p_{\mu}^{(1)} & & \\  &  p_{\mu}^{(2)} &  \\
   & & \ddots \\
 \end{array} \right), \hspace{0.2cm}
\theta^{back}  
 =  \left( \begin{array}{c c c}
  \theta^{(1)} & & \\  &  \theta^{(2)} &  \\
   & & \ddots \\
 \end{array} \right).
\end{equation}

\noindent 
$p_{\mu}$ is decomposed into the trace part and traceless part:

\begin{equation}
p_{\mu}^{(i)}=d_{\mu}^{(i)}{\bf 1} +\tilde{p}_{\mu}^{(i)}, 
\end{equation}

\noindent 
where $d_{\mu}^{(i)}$ is interpreted as the center of mass 
coordinates of the $i$-th blocks. 
We expand the action (\ref{action}) up to the second order of 
the fluctuation and add the following gauge fixing terms to
fix the gauge invariance\cite{IKKT},  

\begin{equation}
S_{\mbox{gauge-fix}}=-Tr( \frac{1}{2}[p_{\mu},a_{\mu}]^{2}
+[p_{\mu},b][p_{\mu},c] ) , 
\end{equation}

\noindent where $c$ and $b$ are ghosts and anti-ghosts, respectively.
The action can be rewritten as 
 
\begin{equation}
\tilde{S} \equiv Tr(\frac{1}{2}a_{\mu}
(P_{\lambda}^{2}\delta_{\mu\nu}-2iF_{\mu\nu})a_{\nu}
-\frac{1}{2}\bar{\varphi}\Gamma^{\mu}P_{\mu}\varphi+bP_{\lambda}^{2}c
+\bar{\varphi}\Gamma^{\mu}\Theta a_{\mu}) .
\end{equation}

\noindent 
where $F_{\mu\nu}$,$P_{\mu}$ and $\Theta$ are adjoint operators 
which act on matrices as follows,

\begin{Eqnarray}
&&P_{\mu}X =[p_{\mu},X] , \cr
&&F_{\mu\nu}X =[f_{\mu\nu},X] \equiv i[[p_{\mu},p_{\nu}],X], \cr
&&\Theta X =\theta X -(-)^{m}X\theta . 
\end{Eqnarray}

\noindent 
\noindent where $m$ is $0$ for bosonic $X$ and $1$ for fermionic $X$. 
One loop effective action $W$ is given by the following equation, 

\begin{equation}
W=-\log\int da d\varphi db dc e^{-\tilde{S}}. 
\end{equation}

\noindent 
We assume that the blocks are separated 
far enough from each other. 
Then the expansion with respect to the inverse powers of the relative distance 
between the two blocks gives the following expression\cite{SuyaTsu, TaylorRaam}.
\begin{Eqnarray}
W &=& \sum_{(i,j)}W^{(i,j)} ,  \cr
W^{(i,j)}&=&-\frac{1}{2}
S{\cal T}r^{(i,j)}
(F_{\mu\nu}F_{\nu\sigma}F_{\sigma\tau}F_{\tau\mu}
-\frac{1}{4}F_{\mu\nu}F_{\mu\nu}F_{\tau\sigma}F_{\tau\sigma})
\frac{1}{(d^{(i)}-d^{(j)})^{8}} \cr
&&-\frac{1}{2}S{\cal T}r^{(i,j)}
(\Theta \Gamma^{\mu}\Gamma^{\nu}\Gamma^{\rho}
 F_{\sigma\mu}F_{\nu\rho}[P_{\sigma},\Theta])
\frac{1}{(d^{(i)}-d^{(j)})^{8}} \cr
&&+W_{\theta^{4}}^{(i,j)} +O(\frac{1}{(d^{(i)}-d^{(j)})^{9}}) .
 \label{mmcalculation}
\end{Eqnarray}

\noindent 
where $S{\cal T}r$ means a symmetrized trace in which we average over 
all possible orderings of the matrices in the trace 
and treat any commutator as single element. 
$W^{(i,j)}$ expresses the interaction between the $i$-th block 
and $j$-th block. 
$W_{\theta^{4}}$ denotes terms including four $\Theta$'s. 
We are not interested in these terms 
in the present discussions. 
$(d^{(i)}-d^{(j)})$ is the distance 
between the center of mass coordinate of the 
$i$-th block and that of the $j$-th block. 
The terms up to $O(r^{-7})$ cancel each other 
when backgrounds are restricted to satisfy the matrix model 
equations of motion (\ref{eom1}) and (\ref{eom2}).
${\cal T}r$ is the trace of the adjoint operators. 
This expression can be rewritten as the form 
of the interaction between the diagonal blocks 
and we take terms 
which are related to 
the exchange of fermionic particles
\footnote{In the paper, we are considering 
only the interaction 
which is related to a photon-photon to 
photino-photino scattering.}:

\begin{Eqnarray}
W^{(i,j)}&=&12\frac{1}{(d^{(i)}-d^{(j)})^{8}}Tr(f_{\mu\lambda}^{(i)}
[\bar{\theta}^{(i)}_{\alpha},\tilde{p}^{(i)}_{\lambda}])
(\Gamma_{\nu})_{\alpha\beta}Tr( f_{\mu\nu}^{(j)}\theta_{\beta}^{(j)}) \cr
&&+12\frac{1}{(d^{(i)}-d^{(j)})^{8}}Tr(f_{\mu\nu}^{(i)}
[\bar{\theta}^{(i)}_{\alpha},\tilde{p}^{(i)}_{\lambda}])
(\Gamma_{\nu})_{\alpha\beta}Tr(f_{\mu\lambda}^{(j)}\theta_{\beta}^{(j)} ) \cr
&&- 6\frac{1}{(d^{(i)}-d^{(j)})^{8}}
Tr([f_{\lambda\rho}^{(i)},\tilde{p}^{(i)}_{\nu}]
\bar{\theta}^{(i)}_{\alpha})
(\Gamma_{\lambda\mu\rho})_{\alpha\beta}
Tr(f_{\mu\nu}^{(j)}\theta_{\beta}^{(j)} ) \cr
&&+4\frac{1}{(d^{(i)}-d^{(j)})^{8}}
Tr([\tilde{p}^{(i)}_{\lambda},f_{\lambda\rho}^{(i)}]
\bar{\theta}^{(i)}_{\alpha})
(\Gamma_{\rho\nu\mu})_{\alpha\beta}
Tr(f_{\mu\nu}^{(j)}\theta_{\beta}^{(j)} ).  
\label{matirxint}
\end{Eqnarray}

Now we can apply these results to noncommutative gauge theory. 
For simplicity, our discussion is restricted to 
$U(1)$ noncommutative gauge theory. 
A graviton exchange process in noncommutative Yang-Mills 
is investigated in \cite{IIKK4038}.
We consider the following noncommutative backgrounds:

\begin{equation}
A_{\mu}^{back}  
 =  \left( \begin{array}{c c}
  p_{\mu}+a_{\mu}^{(1)} & 0 \\ 0  &  p_{\mu}+a_{\mu}^{(2)} \\
 \end{array} \right) ,
\end{equation}

\begin{equation}
\psi ^{back}= \left( \begin{array}{c c}
  \theta^{(1)} & 0 \\ 0  &  \theta^{(2)} \\
 \end{array} \right)  ,
\end{equation}

\noindent 
where $p_{\mu}$ satisfies the noncommutative relation 
(\ref{classicalsol}). 
The rank of $B_{\mu\nu}$ is $d$($=$even), that is, 
the eigenvalues are extended over $d$ dimensional space-time. 
We set $B_{01}=B_{23}=\cdots \equiv B$ for simplicity. 
$a_{\mu}$ and $\theta$ are considered as 
photon and photino fields. 
In the transverse directions, $a_{a}$ is replaced by 
scalar fields $\phi_{a}$. 
By using the mapping rule 
which is summarized in the previous section, 
we have obtained the interactions 
in noncommutative Yang-Mills from (\ref{matirxint}): 

\begin{Eqnarray}
W&=&-\frac{12}{i}\frac{B^{d-8}}{(2\pi)^{d}}\int d^{d}x d^{d}y 
(f_{\mu\lambda}^{(1)}(\partial_{\lambda} \bar{\theta}_{\alpha}^{(1)}))(x)
(\Gamma_{\nu})_{\alpha\beta}
\frac{1}{(x-y)^{8} }( f_{\mu\nu}^{(2)}\theta_{\beta}^{(2)})(y) \cr
&&-\frac{12}{i}\frac{B^{d-8}}{(2\pi)^{d}}\int d^{d}x d^{d}y 
(f_{\mu\nu}^{(1)}(\partial_{\lambda}\bar{\theta}_{\alpha}^{(1)}))(x)
(\Gamma_{\nu})_{\alpha\beta}
\frac{1}{(x-y)^{8}} ( f_{\mu\lambda}^{(2)}\theta_{\beta}^{(2)})(y) \cr
&&+\frac{4}{i}\frac{B^{d-8}}{(2\pi)^{d}}\int d^{d}x d^{d}y 
((\partial_{\lambda} f_{\lambda\rho}^{(1)}) \bar{\theta}_{\alpha}^{(1)})(x)
(\Gamma_{\rho\nu\mu})_{\alpha\beta}
\frac{1}{(x-y)^{8}} 
(f_{\mu\nu}^{(2)}\theta_{\beta}^{(2)})(y) \cr
&&+\frac{6}{i}\frac{B^{d-8}}{(2\pi)^{d}}\int d^{d}x d^{d}y 
((\partial_{\nu} f_{\lambda\rho}^{(1)})\bar{\theta}_{\alpha}^{(1)})(x)
(\Gamma_{\lambda\mu\rho})_{\alpha\beta}\frac{1}{(x-y)^{8}} 
(f_{\mu\nu}^{(2)}\theta_{\beta}^{(2)})(y), 
\end{Eqnarray}

\noindent 
where $f_{\mu\nu}=-B_{\mu\nu}+\partial_{\mu}a_{\nu}-\partial_{\nu}a_{\mu}$.
We have assumed that the external momenta are small 
compared to the noncommutative scale. 
Hence the phase factor which depends on the external momenta 
is dropped.
These terms can be rewritten 
by using the equations of motion (\ref{eomncym}) 
and the following Jacobi identity,  

\begin{Eqnarray}
&&\left( f_{\mu\nu} \partial_{\lambda}
  \bar{\theta}  \right)(x)G(x-y)
+ \left(f_{\nu\lambda} \partial_{\mu}
   \bar{\theta} \right)(x)G(x-y) 
+ \left(f_{\lambda\mu} \partial_{\nu}
    \bar{\theta}  \right)(x)G(x-y)  \cr
&&+ \left(f_{\mu\nu}\bar{\theta} 
 \right)(x) \partial_{\lambda}^{x} G(x-y)
+ \left(f_{\lambda\mu}\bar{\theta} 
  \right) (x)\partial_{\nu}^{x} G(x-y)
+ \left(f_{\nu\lambda}\bar{\theta} 
   \right)(x) \partial_{\mu}^{x} G(x-y)=0. 
\end{Eqnarray}

\noindent 
We can show that order $1/r^{8}$ terms vanish 
up to terms with four $\theta$'s 
and up to total derivative terms and 
we obtain the following expression:

\begin{equation} 
W=\frac{12}{i}\frac{B^{d-8}}{(2\pi)^{d}}\int d^{d}x d^{d}y 
(f_{\rho\nu}^{(1)} \bar{\theta}^{(1)} \Gamma_{\rho})(x)
\slash{\partial}^{x}
\frac{1}{(x-y)^{8}} 
(\Gamma_{\mu}  f_{\mu\nu}^{(2)}\theta^{(2)})(y). 
\end{equation}

\noindent 
We find that only a order $1/r^{9}$ term remains 
because there is a derivative acting on $1/r^{8}$. 

\noindent 
The effect of $B_{\mu\nu}$ decouples from the gravitational 
interaction: 

\begin{Eqnarray}
W&=&-8\frac{12}{i}\frac{B^{d-6}}{(2\pi)^{d}}\int d^{d}x d^{d}y 
(\bar{\theta}^{(1)} )(x)
\slash{\partial}^{x}
\frac{1}{(x-y)^{8}} 
(\theta^{(2)} )(y)   \cr
&&+\frac{12}{i}\frac{B^{d-8}}{(2\pi)^{d}}\int d^{d}x d^{d}y 
(\tilde{f}_{\rho\nu}^{(1)}  \bar{\theta}^{(1)} \Gamma_{\rho})(x)
\slash{\partial}^{x}
\frac{1}{(x-y)^{8}} 
(\Gamma_{\mu}\tilde{f}_{\mu\nu}^{(2)}\theta^{(2)}  )(y) \cr
&=&\frac{12}{i}\frac{B^{d-8}}{(2\pi)^{d}}\int d^{d}x d^{d}y 
(\tilde{f}_{\rho\nu}^{(1)}  \bar{\theta}^{(1)} \Gamma_{\rho})(x)
\slash{\partial}^{x}
\frac{1}{(x-y)^{8}} 
(\Gamma_{\mu}\tilde{f}_{\mu\nu}^{(2)} \theta^{(2)} )(y) , 
\label{matrixresult}
\end{Eqnarray}

\noindent where 
$\tilde{f}_{\mu\nu}=\partial_{\mu}a_{\nu}
-\partial_{\nu}a_{\mu}$ .
In this way, 
we have obtained the long range interactions which decay as $1/r^{9}$ 
in noncommutative Yang-Mills. 
This behavior is independent of the dimensionality of the backgrounds. 
This long range interaction is interpreted as the propagation of the 
massless fermions in ten dimensional supergravity 
and arises from the nonplanar diagram 
in noncommutative Yang-Mills\cite{IKK}. 
In the next section, we see that this interaction 
is interpreted in terms of 
the gravitino exchange 
in ten dimensional supergravity. 

\vspace{0.5cm}
Before finishing this section, let us consider a supersymmetry current 
which is a Noether current associated with supersymmetry.
IIB matrix model has ${\cal N}=2$ supersymmetry 
(\ref{susy1}) and (\ref{susy2}). 
We can determine two supersymmetry currents in the 
matrix model by the Noether method: 

\begin{Eqnarray}
\hat{S}^{\mu(1)}&=&\frac{1}{2g^{2}}\Gamma^{\mu}\psi \cr
\hat{S}^{\mu(2)}&=&\frac{1}{2g^{2}}
f_{\sigma\rho}\Gamma^{\sigma\rho}\Gamma^{\mu}\psi 
  =\frac{i}{2g^{2}}[A_{\sigma},A_{\rho}]\Gamma^{\sigma\rho}\Gamma^{\mu}\psi .
\label{matrixcurrent}
\end {Eqnarray}

\noindent The first supersymmetry current is associated with (\ref{susy1}) and 
the second is associated with (\ref{susy2}). 
The supersymmetry algebra are constructed in \cite{BSS, CMZ}. 
The corresponding currents in noncommutative Yang-Mills theory 
are given by 

\begin{Eqnarray}
S^{\mu(1)}(x)&=&\frac{1}{(2\pi)^{\frac{d}{2}}}\frac{1}{2g^{2}}
   B^{\frac{d}{2}-3}\Gamma^{\mu}\theta(x) \cr
S^{\mu(2)}(x)&=&\frac{1}{(2\pi)^{\frac{d}{2}}}\frac{1}{4g^{2}}
  B^{\frac{d}{2}-4}f_{\sigma\rho}
 \star (\Gamma^{\sigma\rho}\Gamma^{\mu}\theta)(x) ,
 \label{current}
\end {Eqnarray}

\noindent
where $f_{\sigma\rho}=\partial_{\sigma}a_{\rho}
-\partial_{\rho}a_{\sigma} +[a_{\sigma},a_{\rho}]_{\star}$ .
$B_{\mu\nu}$ part, which comes from (\ref{fieldst}), 
in $\hat{S}^{\mu(2)}$ 
is proportional to $\hat{S}^{\mu(1)}$.

These currents are shown to be conserved by using
the equation of motion 
and the Bianchi identity, 

\begin{equation}
[D_{\mu}, S^{\mu}(x)]_{\star} =0  
\end{equation}

\noindent 
In the next section, we investigate 
the interaction between $S^{(2)}_{\mu}$ supercurrents 
via exchange of a gravitino in ten dimensional supergravity. 
We find that supergravity gives the same expression as (\ref{matrixresult}).
We comment about $S^{(1)}_{\mu}$ supercurrent 
in section 5.


\section{Gravitino exchange in supergravity}
\hspace{0.4cm}
In this section, we consider 
a photon-photon to photino-photino scattering via 
exchange of a gravitino, 
which is regarded as 
the supercurrent interactions in ten dimensional supergravity. 

The interaction which is obtained in the matrix model calculation 
is shown to be regarded as 
the interaction between 
$S^{(2)}_{\mu}$ supersymmetry currents in supergravity. 
We also interprete this interaction 
as photon-photon to photino-photino scattering via 
exchange of a gravitino or 
as the non-planar 
one-loop diagrams in the two-point function of 
supercurrents 
in noncommutative Yang-Mills. 

We consider the following gravitino action in ten dimensional supergravity, 

\begin{equation}
S_{\mbox{kinetic}}
=\frac{i}{2} \int d^{10}x \bar{\psi_{\mu}}(x)\Gamma^{\mu\rho\nu}
\partial_{\rho}\psi_{\nu}(x) ,
\end{equation}

\noindent
and determine the gravitino propagator. 
Supergravity theory has local supersymmetry and 
therefore a gauge fixing term must be added 
to obtain the propagator\cite{Nieu}. 
We choose the following gauge fixing term, 

\begin{equation}
S_{\mbox{gauge-fix}}
=-i\int d^{10}x\bar{\psi_{\mu}}(x)
\Gamma^{\mu}\Gamma^{\rho}\Gamma^{\nu}
\partial_{\rho}\psi_{\nu}(x) .
\end{equation}

\noindent 
This gauge is analogous to Feynman gauge in QED. 
Other gauges (ex. Landau gauge) have three derivatives. 
Such a term 
does not appear in the matrix model calculation, 
therefore Feynman gauge is adequate 
for comparing supergravity and the matrix model calculation. 
 The propagator is determined by the following equation, 

\begin{equation}
i(-\frac{1}{2}\Gamma^{\mu\nu\rho}\partial_{\rho}^{x}
-\frac{1}{i}\Gamma^{\mu}\Gamma^{\rho}\Gamma^{\nu}\partial_{\rho}^{x})
\langle \psi_{\nu}(x)\bar{\psi_{\tau}}(y)\rangle
=\eta^{\mu}_{\tau}\delta(x-y)   , 
\end{equation}

\noindent and we find 

\begin{Eqnarray}
\langle \psi_{\nu}(x)\bar{\psi_{\tau}}(y)\rangle
&=&\int \frac{d^{10}k}{i(2\pi)^{10}} 
\frac{1}{4k^{2}}(\Gamma_{\tau}i\slash{k}\Gamma_{\nu}
-6\eta_{\nu\tau}i\slash{k})
e^{ik\cdot(x-y)} \cr
&=& \frac{3}{i8\pi^{5}}
(\Gamma_{\tau}\slash{\partial}\Gamma_{\nu}
-6\eta_{\nu\tau}\slash{\partial})
\frac{1}{(x-y)^{8}} .
\end{Eqnarray}

\noindent 
The gravitino field is assumed to have 
a linear coupling with the supercurrent by the form

\begin{equation}
S=\kappa\int d^{d}x (\bar{\psi}_{\mu}(x)S^{\mu(2)}(x)
 +\bar{S}^{\mu(2)}(x)\psi_{\mu}(x)), 
\end{equation}

\noindent 
where we denote a photon-photino-gravitino 
coupling constant as $\kappa$.  

A tree level gravitino exchange diagram can be calculated by 

\begin{equation}
V=\kappa^{2}\int d^{d}x d^{d}y\bar{S}^{\mu(2)}(x)
\langle \psi_{\mu}(x)\bar{\psi_{\nu}}(y)\rangle S^{\nu(2)}(y). 
\label{currentint}
\end{equation}

\noindent 
We assume that 
the fields which appear in the external lines 
satisfy the on-shell conditions: 

\begin{equation}
\Gamma_{\mu}\partial_{\mu}\theta =0, 
\end{equation}
\begin{equation}
\partial_{\mu}f_{\mu \nu}=\bar{\theta}\Gamma_{\nu}\theta. 
\end{equation}

\noindent 
We denote a normalization factor of $S^{\mu(2)}$ as $C$. 
Then we have 

\begin{Eqnarray}
V&=&-C^{2}\kappa^{2}\frac{3}{i8\pi^{5}}
\int d^{d}x d^{d}y
(f_{\rho\sigma}\bar{\theta}
\Gamma^{\mu}\Gamma^{\sigma\rho})(x)
(\Gamma_{\nu}\slash{\partial}\Gamma_{\mu}
-6\eta_{\mu\nu}\slash{\partial})
\frac{1}{(x-y)^{8}}
(\Gamma^{\tau\lambda}\Gamma^{\nu}f_{\tau\lambda}\theta)(y) \cr
&=&C^{2}\kappa^{2}\frac{9}{i2\pi^{5}}\int d^{d}x d^{d}y
(f_{\rho\sigma}\bar{\theta}\Gamma_{\sigma\rho})(x)
\slash{\partial}
\frac{1}{(x-y)^{8}}
(\Gamma_{\tau\lambda}f_{\tau\lambda}\theta)(y) \cr
&&-C^{2}\kappa^{2}\frac{12}{i\pi^{5}}\int d^{d}x d^{d}y
(f_{\rho\sigma}\bar{\theta}\Gamma_{\sigma\rho})(x)
\partial_{\tau}
\frac{1}{(x-y)^{8}}
(\Gamma_{\lambda}f_{\tau\lambda}\theta)(y) \cr
&&+C^{2}\kappa^{2}\frac{48}{i\pi^{5}}\int d^{d}x d^{d}y
(f_{\rho\sigma}\bar{\theta}\Gamma_{\sigma})(x)
\slash{\partial}
\frac{1}{(x-y)^{8}}
(\Gamma_{\lambda}f_{\rho\lambda}\theta)(y)  \cr
&=&C^{2}\kappa^{2}\frac{3}{i2\pi^{5}}\int d^{d}x d^{d}y
(f_{\rho\sigma}\bar{\theta}\Gamma_{\sigma\rho})(x)
\slash{\partial}
\frac{1}{(x-y)^{8}}
(\Gamma_{\tau\lambda}f_{\tau\lambda}\theta)(y) \cr
&&+C^{2}\kappa^{2}\frac{48}{i\pi^{5}}\int d^{d}x d^{d}y
(f_{\rho\sigma}\bar{\theta}\Gamma_{\sigma})(x)
\slash{\partial}
\frac{1}{(x-y)^{8}}
(\Gamma_{\lambda}f_{\rho\lambda}\theta)(y).   \label{gravitinointeraction}
\end{Eqnarray}

\noindent 
In the last equality, we have used the equations of motion and 
ignore the total derivative terms. 
In view of the tensor structure, 
the first term is identical with the dilatino exchange 
while the second term is 
specific to the gravitino exchange. 

By the same procedure, we next consider 
a spin-$1/2$ component of the supercurrent:

\begin{equation}
S^{(2)}(x)
\equiv \beta \Gamma_{\mu}S^{\mu(2)}(x)
=\beta6Cf_{\rho\sigma}\Gamma^{\rho\sigma}\theta(x),  
\end{equation}

\noindent
where we denote $\beta$ as a normalization factor.
It is expected that this component couples 
to a dilatino field. 
The dilatino kinetic term has the following form: 

\begin{equation}
S=\frac{i}{2}  
\int d^{10}x \bar{\lambda}(x)\Gamma^{\mu}
\partial_{\mu}\lambda(x). 
\end{equation}

\noindent 
Dilatino propagator is given by 

\begin{Eqnarray}
\langle \lambda(x)\bar{\lambda}(y)\rangle
&=&-\int \frac{d^{10}k}{i(2\pi)^{10}} 
\frac{2i\slash{k}}{k^{2}}
e^{ik\cdot(x-y)}  \cr 
&=&- \frac{3}{i\pi^5}\slash{\partial}\frac{1}{(x-y)^{8}}.
\end{Eqnarray}

\noindent 
A dilatino field has a coupling with the spin-$1/2$ component 
of the supercurrent:

\begin{equation}
S=\kappa\int d^{d}x (\bar{\lambda}(x)S^{(2)}(x)
 +\bar{S}^{(2)}(x)\lambda(x)).  
\end{equation}

\noindent 
Therefore interaction between spin-$1/2$ component
is calculated as follows 

\begin{Eqnarray}
V&=&
\kappa^{2}\int d^{d}x d^{d}y\bar{S}^{(2)}(x)
\langle \lambda(x)\bar{\lambda}(y)\rangle S^{(2)}(y)\cr
&=&-\beta^{2} C^{2}\kappa^{2}\frac{108}{i\pi^{5}}
\int d^{d}x d^{d}y
(f_{\rho\sigma}\bar{\theta}
\Gamma^{\sigma\rho})(x)
\slash{\partial}
\frac{1}{(x-y)^{8}}
(\Gamma^{\tau\lambda}f_{\tau\lambda}\theta)(y). 
\label{dilatinointeraction}
\end{Eqnarray}

\noindent 
From (\ref{gravitinointeraction}) and (\ref{dilatinointeraction}), 
we have 

\begin{Eqnarray}
V&=&C^{2}\kappa^{2}\frac{1}{i\pi^{5}}
(\frac{3}{2}-108\beta^{2} )
\int d^{d}x d^{d}y
(f_{\rho\sigma}\bar{\theta}\Gamma_{\sigma\rho})(x)
\slash{\partial}
\frac{1}{(x-y)^{8}}
(\Gamma_{\tau\lambda}f_{\tau\lambda}\theta)(y) \cr
&&+C^{2}\kappa^{2}\frac{48}{i\pi^{5}}\int d^{d}x d^{d}y
(f_{\rho\sigma}\bar{\theta}\Gamma_{\sigma})(x)
\slash{\partial}
\frac{1}{(x-y)^{8}}
(\Gamma_{\lambda}f_{\rho\lambda}\theta)(y). 
\end{Eqnarray}

We compare this supergravity result to 
the matrix model result(\ref{matrixresult}). 
It turns out that $\kappa$ is determined as follows, 
\begin{Eqnarray}
\kappa^{2}&=&4\pi^{5}g^{4} \cr
     &=& 16\pi^{7}g_{s}^{2}\alpha^{\prime 4}.
\end{Eqnarray}
In the last equality, we have expressed $g^{2}$ 
by string coupling $g_{s}$ 
and Regge slope $\alpha^{\prime}$ according to 
the relation $g^{2}=2\pi g_{s}\alpha^{\prime 2}$ \cite{IKKT,IIKK4038}.
We also find that $\beta$ is given by 

\begin{equation}
\beta=\frac{1}{6\sqrt{2}} .
\end{equation}


\section{Conclusions and discussions}
\hspace{0.4cm}
In this paper, we have studied the nonplanar 
two point function 
of the supercurrents in noncommutative Yang-Mills. 
In nonplanar diagrams, 
infrared divergence appears after integrating high 
momentum modes. 
This is a particular phenomenon in noncommutative theories 
and is interpreted to arise from the propagation of 
massless particles. 
We have analyzed this interaction 
from the block-block interaction in the matrix model 
using the connection between the matrix model and 
noncommutative Yang-Mills theory. 
Block-block interactions in IIB matrix model 
are well investigated. 
We examined the block-block interactions 
with fermionic backgrounds at order $1/r^{8}$ 
and mapped to noncommutative Yang-Mills. 
Then we have obtained the interactions which 
decay as $1/r^{9}$ in noncommutative Yang-Mills, 
which is interpreted as the propagation 
of fermionic particle in IIB supergravity. 
Comparing the matrix model result to 
supergravity result, 
we observed the gravitino and dilatino exchange processes 
in the matrix model. 
We also find that one of the two gravitinos 
in IIB supergravity couples to a $S_{\mu}^{(2)}$ 
supersymmetry current.

We have mapped the long range interactions 
at order $1/r^{8}$ in the matrix model to noncommutative Yang-Mills. 
However the long range interaction at order $1/r^{9}$ in 
noncommutative Yang-Mills can appear 
not only order at $1/r^{8}$ in the matrix model calculation, 
which is considered in this paper, but also at $1/r^{9}$. 
Therefore we have to pay attention to sub-leading terms 
(order $1/r^{9}$ terms) 
in the matrix model calculation. 
These terms are calculated in \cite{TaylorRaam}  
in BFSS matrix model. 
One loop amplitudes of BFSS matrix model and 
those of IIB matrix model are related each other by T-duality. 
We can expect that similar terms appear in IIB matrix model 
calculation. 
According to \cite{TaylorRaam},  
there appear order $1/r^{9}$ terms 
which are proportional to the insertion of $d_{\mu}P^{\mu}$ 
into leading terms in (\ref{mmcalculation}). 
Other terms at order $1/r^{9}$ contain three $F_{\mu\nu}$'s. 
Therefore we only need to consider order $1/r^{8}$ terms 
in the matrix model interactions for our investigations of 
gravitino exchange processes. 

We now comment about another gravitino. 
IIB matrix model has ${\cal N}=2$ supersymmetry. 
We have two supercurrents (\ref{matrixcurrent}) in the matrix model.
However after mapping to noncommutative Yang-Mills, 
half of the supersymmetry is spontaneously 
broken due to the backgrounds. 
Therefore $S^{(1)}_{\mu}$ supercurrent 
is not a supersymmetry generator after mapping 
to noncommutative Yang-Mills. 
We easily find that $S^{(1)}_{\mu}$ current does not couple to gravity
because $B_{\mu\nu}$ part decouples from the gravitational interaction 
as in (\ref{matrixresult}). 
To find another gravitino is a problem which is considered in a 
separate paper. 

\vspace{1cm}
\begin{center}
{\bf Acknowledgments}
\end{center}
\hspace{0.4cm}
We are grateful to H.Aoki, M.Hayakawa and S.Iso for valuable 
discussions and comments. 
We also wish to thank T.Suyama and A.Tsuchiya 
for comments about their results.

\vspace{0.5cm}


\end{document}